\documentclass[letterpaper]{emulateapj}
\usepackage{wasysym} 
\usepackage{graphicx}
\usepackage{lscape}

\begin{document}

\title{Infrared Emission by Dust Around $\lambda$ Bootis Stars:\\Debris Disks or Thermally Emitting Nebulae?} 

\shorttitle{$\lambda$ Boo Stars}

\author{J.~R. Mart\'{\i}nez-Galarza\altaffilmark{1}, I. Kamp\altaffilmark{2}, K.~Y.~L. Su\altaffilmark{3}, A. G\'asp\'ar\altaffilmark{3}, G. Rieke \altaffilmark{3}, E.~E. Mamajek\altaffilmark{4}
}

\altaffiltext{1}{Leiden Observatory, Leiden University, P.O. Box 9513, 2300 CA Leiden, The Netherlands}
\altaffiltext{2}{Kapteyn Astronomical Institute, University of Groningen, P.O. Box 800, 9700 AV Groningen, The Netherlands}
\altaffiltext{3}{Steward Observatory, The University of Arizona, 933 N Cherry Ave., Tucson AZ 85721}
\altaffiltext{4}{University of Rochester, Department of Physics \& Astronomy, Rochester NY 14627-0171}

\begin{abstract}

We present a model that describes stellar infrared excesses due to heating of the interstellar (IS) dust by a hot star passing through a diffuse IS cloud. This model is applied to six $\lambda$ Bootis stars with infrared excesses. Plausible values for the IS medium (ISM) density and relative velocity between the cloud and the star yield fits to the excess emission. This result is consistent with the diffusion/accretion hypothesis that $\lambda$ Bootis stars (A- to F-type stars with large underabundances of Fe-peak elements) owe their characteristics to interactions with the ISM. This proposal invokes radiation pressure from the star to repel the IS dust and excavate a paraboloidal dust cavity in the IS cloud, while the metal-poor gas is accreted onto the stellar photosphere. However, the measurements of the infrared excesses can also be fit by planetary debris disk models. A more detailed consideration of the conditions to produce $\lambda$ Bootis characteristics indicates that the majority of infrared-excess stars within the Local Bubble probably have debris disks. Nevertheless, more distant stars may often have excesses due to heating of interstellar material such as in our model. 

\end{abstract}


\section{Introduction}
\label{sec:introduction}

The vast majority of stars follow simple trends in mass, luminosity, age, metallicity, and other parameters as captured in the simplest form through the Russel-Vogt Theorem. The exceptions to these trends therefore must represent interesting events in stellar evolution. $\lambda$ Bootis stars are an outstanding example. They are late B- to early F-type Population I stars with peculiar surface abundances: while light elements like C, N, O and S are roughly solar, heavy elements (Fe-peak) are depleted up to a factor of 100 \citep{Paunzen02}. The ages of $\lambda$ Bootis stars span the entire main sequence lifetimes for their types, with a distribution that has a peak at a rather evolved stage ($\sim$ 1 Gyr). Less than 2\% of all objects in the relevant spectral domain are believed to belong to the $\lambda$ Bootis group, suggesting either a very short time scale (10$^6$ years) for the phenomenon, or uncommon conditions for its occurrence.

Understanding $\lambda$ Bootis stars will provide important insights to more general problems. Due to their extremely shallow convection zones, A-type stars in general and chemically peculiar ones in particular provide excellent natural laboratories for the study of fundamental astrophysical processes such as diffusion, meridional circulation, stellar winds, and accretion of circumstellar and interstellar material.

\cite{Michaud86} suggested that the peculiar chemical abundances on the surfaces of $\lambda$ Bootis stars are due to accretion of circumstellar material that is mixed in the shallow convection zone of the star by the joint action of gravitational settling and radiative acceleration. \cite{Charbonneau93}, \cite{Venn90}, and \cite{Turcotte93} developed this hypothesis further. It naturally explains why the anomalous abundance pattern is similar to that found in the gas-phase of the interstellar medium (ISM), where metallic elements like iron and silicon have condensed into dust grains. \cite{Andrievsky00} suggested a specific mechanism: that $\lambda$ Bootis behavior results from condensation of dust grains in the circumstellar environment, their removal through radiation pressure and the accretion of the remaining metal-poor gas. However, their diffusion/accretion model has many free parameters and fails to explain some observations such as the small proportion of stars with $\lambda$ Bootis characteristics. Alternative theories invoke contact binary systems merging into a single star \citep{Andrievsky97} and unresolved spectroscopic binaries
\citep{Faraggiana99}. 

\begin{figure}[h] \epsscale{0.85} 

\plotone{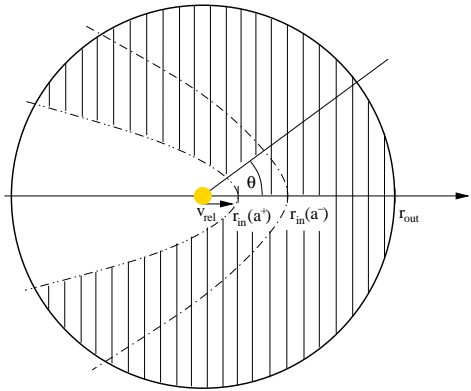} 

\caption{\label{interaction}Geometry of the star-cloud interaction. Two different avoidance radii, corresponding to two different grain sizes are shown.} \end{figure}

More recently, \cite{Kamp02a} have suggested that the $\lambda$ Bootis phenomenon results from interaction along the lines discussed by \cite{Andrievsky00} but between the star and the local ISM (Figure \ref{interaction}). This new hypothesis removes the ad-hoc assumptions and some of the degeneracy of previous models. Different levels of underabundance are produced by different amounts of accreted material relative to the photospheric mass. The small fraction of $\lambda$ Bootis stars is explained by the low probability of a star-cloud interaction and by the effects of meridional circulation, which washes out any accretion pattern a few million years after the accretion has stopped \citep{Kamp02a}. The behavior is suppressed in low temperature stars because of their more massive, difficult-to-contaminate convection zones. Strong stellar winds prevent the accretion of material for very hot stars.

In addition to their peculiar abundance pattern, some $\lambda$ Bootis stars show infrared excesses, suggestive of the presence of dust. Since any ongoing interaction with the ISM will heat interstellar dust, these cases can probe the model proposed by \cite{Kamp02a}. Using IRAS photometry, \cite{King94} found that 5 out of a sample of 56 $\lambda$ Bootis stars have infrared excess in at least one of the satellite's bands. More recently, \cite{Paunzen03} have compiled data from the ISO and IRAS satellites to identify $\lambda$ Bootis stars with excess infrared emission. According to their work, 23\% of 26 well established members of this group show infrared excesses with derived fractional dust luminosities (L$_{\mathrm{IR}}$/L$_*$) ranging from $2.2 \times 10^{-5}$ to $4.3 \times 10^{-4}$. Both the incidence of excesses and the fractional luminosities are similar to the characteristics of planetary debris disks for stars of similar spectral type \citep{Rieke05, Su06}. 

\cite{Kamp02b} used a sample of 21 A-type stars including debris-disks, dust-free stars and six $\lambda$ Bootis stars, and found no fundamental difference between the atmospheric compositions of the stars with and without debris disks. This result suggests that circumstellar material is {\it unlikely} to be linked to the anomalous $\lambda$ Bootis abundances. Thus, the IR excesses of the $\lambda$ Bootis-type stars might be due to a different mechanism, e.g., accretion of interstellar material distributed in a diffuse cloud. This idea is also supported by coronagraphic observations by \cite{Kalas02} that showed some stars previously catalogued as debris disk stars are surrounded by an extended reflection nebulosity. 

In this paper we develop a thermal emission model to explore the \cite{Kamp02a} ISM interaction hypothesis. We emphasize the basic physics behind the model rather than detailed simulations of individual stars. Our most important conclusion is that simple infrared excess SEDs can often be fitted equally well by either a debris disk or interstellar interaction model. As a result, further probing of ISM-related excesses requires detailed modelling of stars where other lines of evidence positively identify such interactions \citep[e.g., for $\delta$ Velorum, ][]{Gaspar08}. 
 
In \S 2 we provide a description of the observations used in this paper and discuss the physical parameters of our sample of stars. We introduce our thermal emission model in \S 3, where we describe the basic physical parameters involved in the calculation of the spectral energy distributions. \S 4 explores the effect of varying different model parameters on the output SEDs. We discuss the implications of our results for the identification of debris disks and the nature of $\lambda$ Bootis stars in \S 5 and
conclude the paper in \S 6.

\medskip
\medskip
\medskip
\medskip

\section{Sample Selection and Observations}

Our sample consists of six A-type stars with well established $\lambda$-Bootis characteristics \citep{Paunzen02} and at distances less than 100 pc. We utilize photometry at 24 and 70 $\mu$m of these stars obtained with the Multiband Imaging Photometer for Spitzer \citep[MIPS; ][]{Rieke04} and reduced with the Instrument Team Data Analysis Tool \citep[DAT; ][]{Gordon05}. The stars were required to have 24 $\mu$m excesses at least 10\% above the photospheres and detected excesses also at 70 $\mu$m. We have excluded binary stars (e.g., HD 198160), where the companion may contaminate the photometry.

Details of the reduction procedures and calibration are provided in \cite{Gordon07} and \cite{Engelbracht07}. Flux densities were extracted using aperture photometry with radii of 14.94" at 24 $\mu$m and 35" at 70 $\mu$m, relative to sky annuli at 29.88" to 42.33" and 39" to 65", respectively. Aperture corrections appropriate to a point source were applied to the photometry. The photometric aperture was selected by consideration of the resolved structure at 24 and 70 $\mu$m for $\delta$ Vel \citep{Gaspar08}, where the bulk of the detected flux would fall within our selected apertures. Since our six stars are all at greater distances than $\delta$ Vel, we should have captured virtually all the flux detectable with MIPS.

To determine the excess properties, we subtracted the photospheric fluxes from the measurements. We determined these fluxes using Kurucz atmospheric models matched to the effective temperature and surface gravity for each star (assumed to have solar abundance). Stellar masses were obtained from the literature \citep[e.g. ][]{Allende99} and they are representative of the spectral type of each star. Stellar radii are fine-tuned to produce a photospheric SED that is in agreement with optical and near-infrared (2MASS) photometry, while total luminosities were calculated by integrating the resulting energy distribution of the scaled Kurucz models over the wavelength range where they are defined. Table 1 lists the fundamental stellar parameters as well as the predicted and measured fluxes at 24 and 70 $\mu$m.

\begin{deluxetable*}{lrrrrrrrrrr}
\tablecolumns{11}
\tablecaption{Physical parameters of the sample stars. \tablenotemark{a}}
\tablehead{
  \colhead{HD No.} &
  \colhead{d} &
  \colhead{T$_{\mathrm{eff}}$} &
  \colhead{$\log_{10}$ L$_{*}$} &
  \colhead{R$_{*}$} &
  \colhead{M$_{*}$} &
  \colhead{F$_{24}$} &
  \colhead{F$_{70}$} &
  \colhead{F$_{\mathrm{pr24}}$} &
  \colhead{F$_{\mathrm{pr70}}$} &
  \colhead{T$_{24-70}$} \\
  \colhead{} &
  \colhead{pc} &
  \colhead{K} &
  \colhead{L$_{\odot}$} &
  \colhead{R$_{\odot}$} &
  \colhead{M$_{\odot}$} &
  \colhead{mJy} &
  \colhead{mJy} &
  \colhead{mJy} &
  \colhead{mJy} &
  \colhead{K}
}
\startdata
 11413 & 74.8 & 7925(124) & 1.32 & 2.4 & 2.6 & 53(1) & 56(2) & 47.71 & 5.42 & 80\\
30422 & 57.5 & 7865(108) & 0.97 & 1.6 & 1.8 & 47(2) & 65(2) & 36.21 & 4.10 & 76\\
31295 & 37.0 & 8920(177) & 1.23 & 1.8 & 2.1 & 177(3) & 438(28) & 101.66 & 11.55 & 71\\ 
110411 & 36.9 & 8930(206) & 1.15 & 1.6 & 2.0 & 147(4) & 256(10) & 91.66 & 10.45 & 84\\
125162 & 29.8 & 8720(156) & 1.20 & 1.9 & 2.0 & 286(3) & 392(6) & 190.74 & 21.59 & 86\\
183324 & 59.0 & 8950(204) & 1.16 & 1.8 & 2.1 & 53(2)& 30(5) & 43.70 & 5.04 & 84\\
\enddata
\tablenotetext{a}{MIPS measurements and predicted photospheric fluxes are shown. Errors shown correspond to measurement uncertainties. Effective temperatures (including errors) were taken from \cite{Paunzen02}, except HD 97633, taken from \cite{Allende99}. T$_{24-70}$ refers to the 24-70 $\mu$m color temperature.}

\end{deluxetable*}

\section{Model Description}

Following the \cite{Kamp02a} interaction model, here we consider an A-type star moving through a uniform density, optically thin ISM cloud. The relative motion between the star and the cloud is described by a constant relative speed $v_{\rm{rel}}$. The emission of the central star irradiates the sub-micron dust grains present in the cloud, warming them up to equilibrium temperatures, at which they re-emit the energy in the IR. Since the ISM cloud is optically thin, thermal radiation emitted by the warm grains will not be reprocessed by the cloud itself. Our model calculates the resulting SED at mid-infrared wavelengths.

\subsection{Geometry of the cloud}

Our model assumes a spherical cloud, penetrated by an axially symmetric paraboloidal cavity excavated by the radiation pressure of the star as it moves through the cloud, as described by \cite{Artymowicz97}. We can describe this geometry mathematically using spherical coordinates ($r$, $\theta$, $\phi$). Because of the axisymmetrical character of the cavity, only a radial coordinate ($r$) and an azimuthal coordinate ($\theta$) are needed to describe its surface, which is symmetrical with respect to $\phi$. We choose the coordinate system so that the star is at the origin ($r = 0$) and marks the center of the spherical structure. The ray with zero azimuthal angle coincides with the direction of relative motion (Figure \ref{interaction}). The equation describing the surface of the cavity is then

\begin{equation}
\label{eq1}
r=r_{\rm{in}} \left( {\frac{2}{1+\cos \theta }} \right).
\end{equation}

\noindent
The geometry is parametrised by the avoidance radius, $r_{\mathrm{in}}$, which determines the shape and size of the cavity\footnote{The behavior is analogous to Rutherford scattering of positively charged particles by a positively charged atomic nucleus.}. The grains never approach the star closer than $r_{\mathrm{in}}$. The avoidance radius can be estimated by considering energy conservation. For a relative velocity between the star and the cloud ($v_{\mathrm{rel}}$),  a ratio of radiation pressure to gravity ($\beta$), and a stellar mass of M$_*$, it can be show that 

\begin{equation}
\label{eq2}
r_{\rm{in}} =\frac{2(\beta -1)\rm{GM}_\ast }{v_{\mathrm{rel}}^2 },
\end{equation}

\noindent
where G is the Newtonian gravitational constant \citep{Artymowicz97}.

\cite{Burns79} show how to calculate $\beta$, which determines $r_{\mathrm{in}}$ as a function of the grain size $a$. We briefly reproduce their approach. The gravitational attraction of the star upon a spherical particle of radius $a$ and density $\rho_{\mathrm{dust}}$ (g cm$^{-3}$) that is located at a distance $r$ from the star is 

\begin{equation}
\label{eq3}
F_{\rm g}^a =\frac{4}{3}\pi a^3\rho _{\rm{dust}} \rm{GM}_\ast /r^2.
\end{equation}

\noindent
The pressure force due to the radiation field of the star is given in terms of the stellar luminosity L$_*$ and the geometric cross section $A$ of a spherical grain of radius
$a$:

\begin{equation}
\label{eq4}
F_{\rm r}^a =\left( {\frac{\rm{L}_\ast }{4\pi r^2c}A} \right)\overline {Q_{\rm{pr}}^a } 
\end{equation}

\noindent
where $\overline{Q_{\mathrm{pr}}^a}$ is the average radiation pressure coefficient for grains of size $a$. To calculate this coefficient, we first compute
the grain-size-dependant radiation pressure efficiency Q$_{\mathrm{pr}}^a$ as a function of wavelength:  

\begin{equation}
\label{eq5}
Q_{\rm{pr}}^a (\lambda )=Q_{\rm{abs}}^a +Q_{\rm{sca}}^a (1-\langle \cos \alpha^a \rangle ).
\end{equation}

\noindent
Both the scattering coefficient $Q_{\mathrm{sca}}^a$ and the absorption coefficient $Q_{\mathrm{abs}}^a$, 
as well as the scattering angle, $\alpha^a$, are
functions of the wavelength. They are available for appropriate grain parameters in tabulated form.
Once we have determined the shape of $Q_{\mathrm{pr}}^a$($\lambda$), we compute $\overline{Q_{\mathrm{pr}}^a}$ as a weighted average of $Q_{\mathrm{pr}}^a$ over the wavelength spectrum:

\begin{equation}
\label{eq6}
\overline {Q_{\rm{pr}}^a} =\frac{\int {F_{\rm{K}} (T_\ast )Q_{\rm{pr}}^a (\lambda )\rm{d}\: \lambda } 
}{\int {F_{\rm{K}} (T_\ast )\rm{d}\: \lambda } }
\end{equation}

\noindent
where $F_{\rm{K}}(\rm{T}_*)$ is the energy density of the respective Kurucz model. We can then compare the radiation force to the gravitational force by means of their ratio:

\[
\beta^a =F_{\rm{r}} /F_{\rm{g}} =(3\rm{L}_\ast /16\pi \rm{GM}_\ast \rm{c})(\overline {Q_{\rm{pr}}^a } /a\rho 
_{\rm{dust}} )
\]
\begin{equation}
\label{eq7}
=0.57\overline {Q_{\rm{pr}}^a } \left( {\frac{\rm{L}_\ast }{\rm{M}_\ast }} 
\right)\left( {\frac{a}{\mu m}} \right)^{-1}\left( {\frac{\rho _{\rm{dust}} 
}{\rm{g}\;\rm{cm}^{-3}}} \right)^{-1}
\end{equation}

\noindent
where L$_*$ and M$_*$ are in solar units. $\rho_{\mathrm{dust}}$ should be distinguished from the total dust mass density of the cloud, $\rho_{\mathrm{cloud}}$, which is one of the free parameters in our model. Eq. (7) shows the dependence of $\beta^a$ on the grain size. As a consequence, under the same physical conditions, grains of the same composition but different sizes will have different avoidance radii, which in turns leads to different equilibrium temperatures. Mathematically, this dependence arises not only from the factor $a^{-1}$ in Eq.\ (7), but also because the optical properties of a grain, (i.e., $Q_{\mathrm{abs}}^a$ and $Q_{\mathrm{sca}}^a$) are functions of the grain size. In \S 4 we will discuss further variations in $r_{\mathrm{in}}$ due to different compositions of the grain material. Because of different grain properties, the model cavity has a large range of avoidance radii, resulting in a corresponding range in the cavity size and in the grain equilibrium temperatures.

\subsection{Relative velocities}

As shown in Eq.\ (2), the shape of the cavity depends on the velocity of the star relative to the cloud. 
Determination of the heliocentric velocity for the interstellar material close to our stars, $v_{\rm{ISM}}$, and hence, of the relative velocity of the system ($v_{\mathrm{rel}}$), is not straightforward. In some instances, there are candidate clouds with known velocity vectors in the local ISM which could be responsible for interacting with the $\lambda$ Bootis stars. A catalogue of local nearby clouds of warm ISM and their velocity vectors and approximate outlines is presented in \cite{Redfield08}. A program called "windsock" was written by Eric Mamajek which estimates the relative 3D velocities of ISM clouds and the target stars, and calculates the projected relative tangential and radial motions appropriate for comparing to the detailed dust interaction models. The cloud velocity vectors were adopted from \cite{Redfield08}, and the velocity vectors of the stars were estimated from revised Hipparcos astrometry from \cite{Leeuwen} and the compiled radial velocity catalogue of \cite{Gontcharov06} using the formulae of \cite{Johnson87}.  Table 2 shows the velocities for each of our stars relative to several candidate clouds, as calculated using windsock. Since we cannot determine in each case which of these clouds actually interacts with the star, we have decided to study the behavior of the model at two limiting velocities, which cover approximately the range of velocities in Table 2. These velocities are 15 km s$^{-1}$ and 40 km s$^{-1}$. It is important to mention that the relative velocity for two of our stars (HD 125162 and HD 30422) is closer to our lower velocity limit (15 km s$^{-1}$), while for the remaining five stars this velocity is closer to the upper limit (40 km s$^{-1}$). In \S 4 we study the effect of the relative velocity on the infrared excess properties and we estimate the corresponding uncertainties.

\begin{deluxetable*}{lrrrrrrrrrrr}
\tablecolumns{12}
\tablecaption{Windsock model velocities. \tablenotemark{a}}
\tablehead{
  \colhead{HD No.} &
  \colhead{$v_{\mathrm{LIC}}$} &
  \colhead{$v_{\mathrm{G}}$} &
  \colhead{$v_{\mathrm{Vel}}$} &
  \colhead{$v_{\mathrm{Blue}}$} &
  \colhead{$v_{\mathrm{Aur}}$} &
  \colhead{$v_{\mathrm{Gem}}$} &
  \colhead{$v_{\mathrm{NGP}}$} &
  \colhead{$v_{\mathrm{Leo}}$} &
  \colhead{$v_{\mathrm{Mic}}$} &
  \colhead{$v_{\mathrm{Aql}}$} &
  \colhead{$v_{\mathrm{Eri}}$}  
}
\startdata
 11413 & 40.0 & 43.4 & 58.0 & - & - & - & - & - & - & - &-\\
 30422 & 12.2 & - & - & 7.3 & 9.6 & - & - & - & - & - & -\\
 31295 & 27.5 & - & - & - & 18.6 & 29.0 & - & - & - & - & -\\
 110411 & - & - & - & - & - & - & 56.8 & 43.3 & - & - & -\\
 125162 & - & - & - & - & - & - & 2.3 & 10.4 & - & - & -\\
 183324 & - & 42.0 & - & - & - & - & - & - & 42.1 & 64.7 & 36.8\\ 
\enddata 
\tablenotetext{a}{For each star, velocities in km s$^{-1}$ are shown for clouds that are more likely to be interacting with it according to \cite{Redfield08}.}
\end{deluxetable*}

\subsection{Thermal emission}
With the geometric distribution of the dust defined, we can predict the infrared fluxes in the MIPS bands. The fixed parameters in our model are the relative velocity $v_{\mathrm{rel}}$, the mass to luminosity ratio (M/L) of the star, its radius and heliocentric distance d, as well as the composition, size and optical properties of the dust grains. All except the last set of parameters have already been discussed. 

For the grains, we assume an interstellar composition, i.e. the simplistic homogeneous sphere models \citep{Mathis77}, hereafter MRN, with two (graphite, silicate) components and a MRN size distribution of the form $\rm{d}n = \rm{C}a^{-3.5} \rm{d}a$ with lower and upper limits at $a_{\mathrm{min}}= 0.005$ $\mu$m and $a_{max} = 0.25$ $\mu$m respectively \citep{Mathis90}. Graphite and silicate populations are treated independently. The free parameters in our model are then just the outer radius of the cloud $r_{\mathrm{out}}$, i.e., its size, and the total mass density of its dust component,
$\rho_{\mathrm{cloud}}$.

To calculate IR fluxes, we assume that the grains are in radiative equilibrium. Grains absorb stellar radiation and re-emit it at temperature $T_{\rm{g}}$. We use the Kurucz model described in \S 2 for the stellar radiation field. The radiative equilibrium at a distance $r$ from the star can be expressed as

\begin{equation}
\label{eq9}
\frac{1}{4}\frac{\rm{R}^2}{r^2}\int {Q_{\mathrm{abs}}^a (\lambda )F_{\rm{K}} (\lambda )\rm{d}\lambda 
=\int {Q_{\mathrm{abs}}^a } } (\lambda )B(\lambda ,T_{\rm{g}} )\rm{d}\lambda. 
\end{equation}

\noindent
$Q^{a}_{\rm{abs}}$($\lambda$) is the absorption efficiency for grains of size $a$. Since we assume thermal equilibrium, Kirchoff's law applies and the absorptivity of the dust grains equals their emissivity. $Q_{\rm{abs}}$ thus appears on both sides of Eq.\ (9), but evaluated over different wavelength regimes. By solving this equation numerically, we find the equilibrium temperature $T_{\rm{g}}$ as a function of distance, grain size and grain composition.

The functional forms of $Q^{a}_{\mathrm{abs}}$($\lambda$) for interstellar dust grains of the desired composition have been published by \cite{Draine84} and \cite{Laor93}, for grain sizes covering the range 0.001 $\mu$m $\le$ $a$ $\le$ 10 $\mu$m and for wavelengths between 0.001 $\mu$m and 1000 $\mu$m. We use these tables in our integration of Eq.\ (8). For each grain size, a different avoidance radius is calculated according to Eq.\ (2). The grain number density of the cloud,  $n$, is related to the grain size by

\begin{equation}
\label{eq10}
n(a)=\rm{C}a^{-3.5}\rm{d}a.
\end{equation}

The integration is performed over the range of grain sizes. The astronomical silicates ($\rho_{\mathrm{dust}}$ = 3.3 g cm$^{-3}$) and graphite ($\rho_{\mathrm{dust}}$ = 2.25 g cm$^{-3}$) both have a size distribution described by Eq.\ (9), and each of them carries half the total mass ($M_{\mathrm{dust}}$) of the dust component of the ISM cloud. Mathematically, this leads to the normalisation equation:

\begin{equation}
\label{eq11}
\int {n(a)m(a)V(a)\rm{d}a}=\frac{\rm{M}_{\rm{dust}} }{2}=\frac{\rho V_{\rm{dust}} }{2}
\end{equation}

\noindent
where  $n(a)$ and $m(a) = 4\pi\ a^3\ \rho_{\mathrm{dust}}/3$ are respectively the number density and individual grain mass as functions of the grain size, $V(a)$ is the volume occupied by grains of size $a$ (each grain size has a different avoidance radius) and $V_{\rm{dust}}$ is the total volume occupied by the dust cloud. Combining Eq.\ (9) and Eq.\ (10) and solving for the normalisation constant C, we get:

\begin{equation}
\label{eq12}
C=\frac{3\rho V_{\mathrm{dust}} }{\rho _{\mathrm{dust}} }\left( {\int {V(a)a^{-0.5}\rm{d}a} } 
\right)^{-1}
\end{equation}

\noindent
which is valid for any $\rho_{\mathrm{dust}}$. Taking into account Eq.(1) for each grain size, we have calculated the normalisation constants for each population of grains as well as the total dust mass of the cloud, $M_{\rm{dust}}$.

Now that we have characterised the grain size distribution, we can calculate the infrared flux. For each grain size, we integrate the contributions of the modified blackbodies (dust grains), as determined by the distribution of equilibrium temperatures (Eq.(8)) over the cloud volume:

\begin{equation}
\label{eq13}
\rm{d}F_a (\lambda )=n(a)\frac{\pi a^2}{d^2}Q_{\rm{abs}}^a (\lambda )\int_V {B(\lambda 
,T_{\rm{g}} (r))\rm{d}V} 
\end{equation}

\noindent
Finally, we integrate over the grain size distribution to obtain the spectral energy distribution (SED):

\begin{equation}
\label{eq14}
F(\lambda )=\int_{a_{\min } }^{a_{\max } } {n(a)\frac{\pi a^2}{d^2}Q_{\rm{abs}}^a 
(\lambda )} \int_V {B(\lambda ,T_{\rm{g}} (r))\rm{d}V} 
\end{equation}

\noindent
We fit the MIPS photometry with this function. 

If the IR excesses of the $\lambda$ Bootis stars are indeed due to a thermally emitting nebulosity, the MIPS observations discussed in \S 2 may partially resolve the sources. In that case, the fitting to the photometry needs to allow for the source extent. We discuss this issue in \S 4.1. We then continue to describe the underlying model physics and estimates of cloud dimensions and density to assess the plausibility of the star-ISM interaction model.

\section{Exploring the parameter space}

To assess the uncertainties involved in our model and to test its validity, we perform a series of trial calculations aimed at establishing the impact of individual parameters on the SED and derived quantities. We also examine the influence of grain composition and total dust density. For these tests, we use the stellar and cloud parameters derived for HD 125162, the prototype of the $\lambda$ Bootis stars.

\subsection{Outer radius of the cloud}

Because of its relatively small heliocentric distance, HD 125162 is the most demanding  example to assess the uncertainties due to our selected aperture for photometry. At the distance to this star (29.8 pc), the aperture radius of 14.94" used to perform the point source photometry at 24 $\mu$m corresponds to a physical radius of 445 AU, whereas an aperture radius of 35", used to perform the photometry at 70 $\mu$m, corresponds to a physical radius of 1043 AU. As discussed in the next section, to produce the observed fluxes, our model suggests an outer radius of 2200 AU for the diffuse interstellar cloud, when a velocity of 40 km s$^{-1}$ is used. To estimate the uncertainties in our best fit parameters due to the difference between the aperture radius and the actual size of the cloud, we compare the fluxes at 24 and 70 $\mu$m obtained for the entire cloud ($r_{\rm{out}}$ = 2200 AU) with the corresponding fluxes that now only include material within the corresponding aperture radii (445 and 1043 AU). At 24 $\mu$m this flux is calculated in the same way as we obtained the photometric measurements, i.e., we subtract the flux from the sky annulus and apply an aperture correction factor:

\begin{equation}
\label{eq15}
F(24\, \mu m)=1.143(F_{\rm{24}}^{\rm{aperture}} -F_{\rm{24}}^{\rm{annulus}} )
\end{equation}

\noindent
We obtain a flux that is only 20\% smaller than the flux at 24 $\mu$m calculated using the entire cloud. At 70 $\mu$m, the emission from the sky annulus is comparable to that from the aperture and hence we did not subtract it. We compare the flux obtained for the entire cloud with that calculated with an outer radius of 1043 AU, corresponding to the 35" aperture. In this case, the flux turns out to be a factor of 2.5 smaller than the one obtained for the entire cloud. This result suggests that a larger photometric aperture should have been employed. However, there are significant difficulties in identifying and measuring low surface brightness extensions at 70 $\mu$m and it is not clear that we would improve the overall accuracy of the measurement. Since HD 125162 is the closest star in our target list, we consider these uncertainties to be an upper limit for our sample.

\begin{figure} [b]\epsscale{0.85} \begin{center} \rotatebox{90}{\plotone{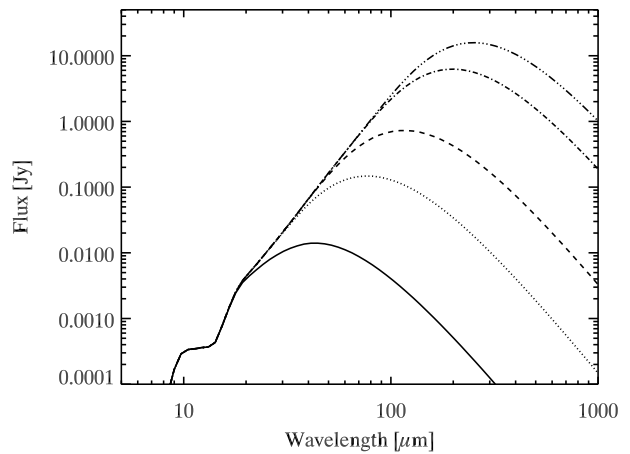}} \end{center} \caption{\label{outer} Dependence of the dust IR emission on the outer radius of the cloud. The size range goes from typical disks sizes (500 AU, solid line) up to typical sizes of diffuse interstellar clouds (100000 AU, triple dot-dashed line), with intermediate values at 3000 AU (dotted line), 10000 AU (dashed line) and 50000 AU (dot-dashed line).} \end{figure}

Figure \ref{outer} illustrates a more general case. It shows dust SEDs for five diffuse clouds with different outer radii. To exclude other possible effects, we have used here only one composition (silicates) and a single grain size (0.008 $\mu$m, corresponding to the weighted mean of the distribution). The stellar parameters correspond to HD 125162, with a velocity corresponding to our higher limit of 40 km s$^{-1}$. Even though, as we stated before, for HD 125162 the relative velocity is closer to our lower limit of 15 km s$^{-1}$, here we are only interested in the impact that varying the outer radius has on the model and hence the velocity used is not relevant. Additionally, as showed in \S4.3, the SED does not change significantly in the mid-infrared region within a range of velocities between 15 and 50 km s$^{-1}$. The total dust mass density is kept constant, and the outer radius ranges from an upper limit of the typical radii of circumstellar disks ($\sim$ 500 AU) up to the size scale of diffuse clouds in the interstellar medium ($\sim$ 10$^5$ AU). As the cloud becomes larger, the total emission increases and the maximum of the SED is shifted to longer wavelengths. As we increase the outer radius of the cloud we increase the contribution to the SED of external, colder material, which dominates the total mass. Since the density is kept fixed, increasing values of r$_{\rm{out}}$ imply an increase in the total cloud mass and therefore an increase of the total emission. A tendency for ISM-interaction infrared excesses to be "warm" \citep{Gaspar08} might be exaggerated observationally by the inherent measurement bias against this dilute, extended emission component.

\subsection{Relative velocity}

\begin{figure} [t]\epsscale{0.85} \begin{center} \rotatebox{90}{\plotone{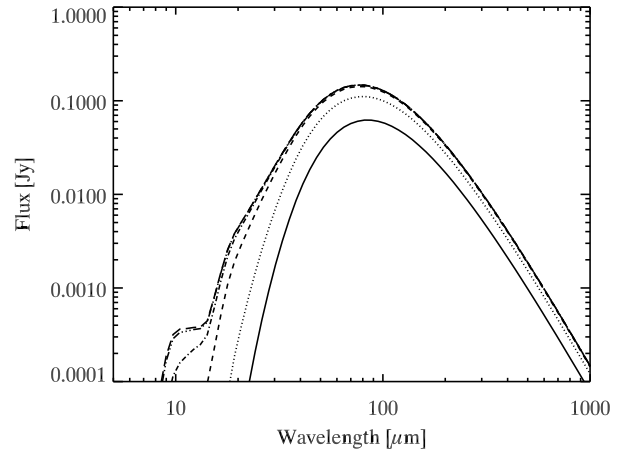}} \end{center} \caption{\label{velocity1} Dependence of the IR emission on the relative star-cloud velocity. Velocities are: 0.9 km s$^{-1}$ (solid line), 1.5 km s$^{-1}$ (dotted line), 4 km s$^{-1}$ (dashed line), 10 km s$^{-1}$ (dot-dashed line), 35 km s$^{-1}$ (triple dot-dashed line) and 100 km s$^{-1}$ (long-dashed line)} \end{figure}

Given the stellar space velocities and the available measurements on the local interstellar cloud kinematics, the parameter $v_{\rm{rel}}$ can lie approximately between 15 km s$^{-1}$ and 40 km s$^{-1}$, as stated. Figure \ref{velocity1} shows the calculated SEDs for a cloud of silicate grains with single grain size $a = 0.008$ $\mu$m around HD 125162. The relative velocity spans a range between 0.9 km s$^{-1}$ and 100 km s$^{-1}$, while all the other parameters remain fixed. Changing the relative velocity changes the avoidance radius according to Eq.\ (2). However, for the range of velocities of interest (15 km s$^{-1}$ to 40 km s$^{-1}$), the SEDs are very similar at wavelengths larger than 20 $\mu$m, and approach an asymptotic SED. This asymptotic case corresponds to high velocities ($v_{\mathrm{rel}}$ $>$ 10 km s$^{-1}$ for silicate grains), in which case $r_{\rm{in}}=0$ and the cavity vanishes, since there is no time to excavate it. It is good to remind ourselves, however, that physically this will not be the case, since there will always be an inner radius set by the sublimation temperature of the grains \citep[$\sim$ 1500 K for silicate grains, ][]{Salpeter77}. We see in Figure \ref{velocity1} again a color change of the SED similar to the one described in \S 4.1: as the relative velocity becomes smaller, the avoidance radius becomes larger and colder material dominates the emission. Carbonaceous grains behave similarly, but due to their larger radiation pressure efficiency the velocities needed to reach the limiting case are of the order of $v_{\rm{rel}}$ $>$ 100 km s$^{-1}$. As shown in Tables 3 and 4, for a particular star-cloud system, derived cloud densities at 15 km s$^{-1}$ are approximately 3.5 larger than in the 40 km s$^{-1}$ case, while derived outer radii are about 1.6 times smaller. The total mass needed to produce the measured excess, however, remains almost the same in both cases.

\subsection{Dust composition}

\begin{figure} [t]\epsscale{0.85} \begin{center} \rotatebox{90}{\plotone{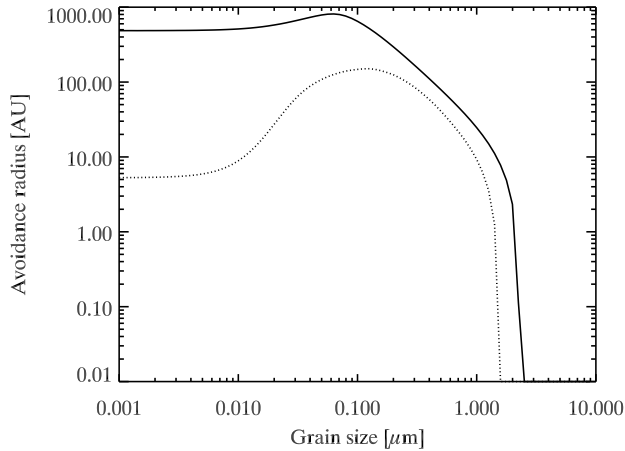}} \end{center} \caption{\label{composition1} Dependence of the avoidance radius on the composition of dust grains. The two different compositions are: astronomical silicate grains (dotted line), and graphite grains (solid line). The encounter velocity is 15 km s$^{-1}$, and the stellar parameters are those for HD 125162.} \end{figure}

For the purposes of this paper, we use a mixed population of astronomical silicate and graphite dust grains, each of them providing 50\% of the total dust mass. We assume bare grains, without ice mantles. The complex refractive indices and the absorption and scattering properties of dust grains \citep[e.g. ][]{Draine84, Laor93} imply that at the wavelengths at which A-type stars emit most of their flux, carbonaceous compounds are more efficient at absorbing UV photons compared to astronomical silicates, according to Eq.\ (6). Hence, the radiation pressure on graphite is larger than for the silicates and, as a consequence, carbonaceous grains will be located at larger distances from the star. Figure \ref{composition1} depicts the situation for a particular relative velocity. The avoidance radii of silicate grains are about 2 orders of magnitude smaller than those of graphite grains. However, the difference in absorption coefficients also means that graphite is heated more efficiently by the radiation field, reaching higher equilibrium temperatures and hence contributing more at near-IR wavelengths than silicates. Our model shows that for radiation fields typical of our stars, most of the infrared flux, in both MIPS bands, is coming from the carbonaceous grains, with the fluxes produced by silicates being at least one order of magnitude smaller, in spite of the fact that silicate grains get much closer to the star.

Figure \ref{composition1} illustrates another point. The maximum difference in avoidance radii between the two compositions occurs at the peak of the graphite emission ( $a \sim 0.06$ $\mu$m). For silicate grains the avoidance radius peaks at a slightly larger grain size of about 0.1 $\mu$m. As the grain size increases to values larger than 0.1 $\mu$m, the avoidance radii converge to zero. However, our grain size distribution, with a sharp cutoff at 0.25 $\mu$m, does not have grains that are large enough to be accreted onto the stellar surface. In any case,  a limiting avoidance radius is set by the sublimation of grains. 

\subsection{Total dust mass density}

\begin{figure} \epsscale{0.85} \begin{center} \rotatebox{90}{\plotone{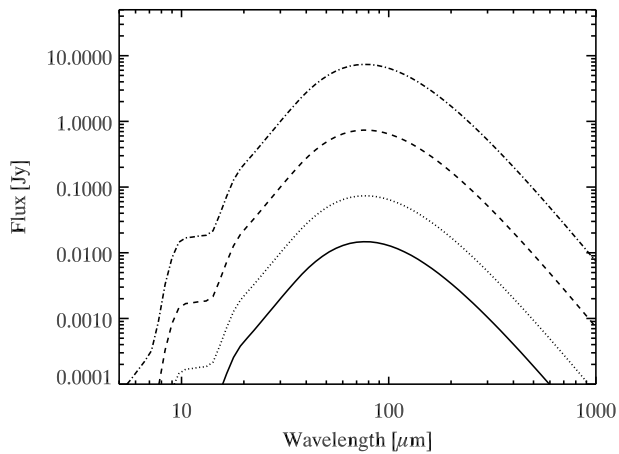}} \end{center} \caption{\label{density} Dependence of the dust IR emission on the total dust particle density of the cloud. Density values shown correspond to: $0.45$ cm$^{-3}$ (solid line), $2$ cm$^{-3}$ (dotted line), $20$ cm$^{-3}$ (dashed line) and $215$ cm$^{-3}$ (dash-dotted line). } \end{figure}

We have also studied the effect of varying the total dust mass density of the cloud, $\rho_{\rm{cloud}}$. Figure \ref{density} shows four different spectral energy distributions for a cloud around HD 125162. Again, we have used a grain size of 0.008 $\mu$m and silicate composition. The total gas density, $n_{\mathrm{gas}}$, varies between 0.45 cm$^{-3}$ and 215 cm$^{-3}$, assuming a gas to dust ratio of 100 and a mean molecular weight of 1.4 times that of atomic hydrogen. For comparison typical diffuse clouds in the galaxy have number densities of the order of 10-100 cm$^{-3}$. As we go from very diffuse clouds to higher densities, keeping all the other parameters constant, the total amount of material increases, and similarly, the total flux emitted by the cloud rises. The position of the emission peak remains the same, at about 55 $\mu$m, since we are not changing the equilibrium temperatures of the grains when we increase the density.

\section{Discussion}

\subsection{Prevalence of ISM-interaction excesses}

For a given star, our model takes the stellar parameters, the dust optical parameters and the relative velocity as an input, calculates avoidance radii for grains with different sizes using Eq.(2) and finds the total dust mass density and size of the cloud that best fit the MIPS photometry. This is done for the two limiting velocities, 15 and 40 km s$^{-1}$. We perform a non-linear least squares fitting to the MIPS photometry. Using the reduced $\chi$-squared minimisation method proposed by \cite{Lampton76}, we have derived the parameters $\rho_{\mathrm{cloud}}$ and $r_{\rm{out}}$ that best fit the photometry for each of our sample stars. Statistical errors for these parameters have been estimated at a 90\% confidence level. We obtain number densities for the gas in our clouds, as well as total masses of the clouds. The total amount of material in the clouds and their gas densities allow a comparison with typical values for IS clouds. They are entered in Tables 3 and 4, where we also list average metallicities for our sample stars. Note that the cloud density values, $n_{\mathrm{gas}}$ are very low, close to current detection limits. Thus, nearly any identified IS cloud can potentially produce an excess when an A-star penetrates it. Very diffuse clouds with $n_{\mathrm{gas}}<1$ cm$^{-3}$ are more likely to produce an excess at higher velocities.

\begin{deluxetable*}{lrrrrrrrrrr}
\tablecolumns{11}
\tablecaption{Model results for $v_{\mathrm{rel}}=15\, \mathrm{km}\, \mathrm{s}^{-1}$.\tablenotemark{a}}

\tablehead{
  \colhead{HD No.} &
  \colhead{$n_{\mathrm{gas}}$} &
  \colhead{$r_{out}$} &
  \colhead{$M_{\mathrm{dust}}$} &
  \colhead{$r_{\mathrm{acc}}$} &
  \colhead{$\dot{M}$} &
  \colhead{$t_{\mathrm{acc}}$} &
  \colhead{$M_{\mathrm{acc}}$} &
  \colhead{$r_{\mathrm{aper,24}}$} \tablenotemark{b} &
  \colhead{$r_{\mathrm{aper,70}}$} \tablenotemark{c} &
  \colhead{$Z_{\mathrm{aver}}$} \tablenotemark{d}\\
  \colhead{} &
  \colhead{cm$^{-3}$} &
  \colhead{AU} &
  \colhead{$\mathrm{M}_{\rm{Moon}}$} &
  \colhead{AU} &
  \colhead{$\times 10^{-14}\mathrm{M}_{\astrosun}\: \mathrm{yr}^{-1}$} &
  \colhead{$\mathrm{yr}$} &
  \colhead{$\times 10^{-11}\mathrm{M}_{\astrosun}$} &
  \colhead{AU} &
  \colhead{AU} &
  \colhead{dex}
}
\startdata
 11413 &  1.00 &  2729 &  0.091 &  16.33 &  1.05 & 1728 &  1.82 & 1117 & 2617 & -1.07\\
 30422 &  3.12 &  1479 &  0.045 &  10.97 &  1.47 &  937 &  1.38 & 859 & 2011 & - \\
 31295 &  3.01 &  2245 &  0.151 &  12.97 &  1.98 & 1422 &  2.82 & 552 & 1294 & -0.91\\
 110411&  3.71 &  1591 &  0.065 &  12.59 &  2.31 & 1008 &  2.33 & 551 & 1292 & -0.83\\
 125162&  5.48 &  1402 &  0.067 &  12.72 &  3.47 &  888 &  3.08 & 445 & 1043 & -1.35\\
 183324&  2.80 &  1085 &  0.011 &  13.03 &  1.86 &  687 &  1.28 & 881 & 2065 & -1.41\\
\enddata
\tablenotetext{a}{The uncertainties in the best-fit parameters ($r_{\mathrm{out}}$ and $\rho$) are of about 10\%, as calculated using the $\chi^2$ minimisation. $n_{\mathrm{gas}}$ corresponds to the total number gas density of the cloud, while $t_{\mathrm{acc}}$ is the calculated timescale of the star-cloud interaction.}
\tablenotetext{b}{Physical aperture radius at the distance to the star at 24 $\mu$m.}
\tablenotetext{c}{Physical aperture radius at the distance to the star at 70 $\mu$m.}
\tablenotetext{d}{Average metallicities calculated for the stars based on \cite{Heiter02}.}

\end{deluxetable*}

\begin{deluxetable*}{lrrrrrrrrrr}
\tablecolumns{11}
\tablecaption{Model results for $v_{\mathrm{rel}}=40\, \mathrm{km}\, \mathrm{s}^{-1}$.\tablenotemark{a}}

\tablehead{
  \colhead{HD No.} &
  \colhead{$n_{\mathrm{gas}}$} &
  \colhead{$r_{out}$} &
  \colhead{$M_{\mathrm{dust}}$} &
  \colhead{$r_{\mathrm{acc}}$} &
  \colhead{$\dot{M}$} &
  \colhead{$t_{\mathrm{acc}}$} &
  \colhead{$M_{\mathrm{acc}}$} &
  \colhead{$r_{\mathrm{aper,24}}$} \tablenotemark{b} &
  \colhead{$r_{\mathrm{aper,70}}$} \tablenotemark{c} &
  \colhead{$Z_{\mathrm{aver}}$} \tablenotemark{d}\\
  \colhead{} &
  \colhead{cm$^{-3}$} &
  \colhead{AU} &
  \colhead{$\mathrm{M}_{\rm{Moon}}$} &
  \colhead{AU} &
  \colhead{$\times 10^{-16}\mathrm{M}_{\astrosun}\: \mathrm{yr}^{-1}$} &
  \colhead{$\mathrm{yr}$} &
  \colhead{$\times 10^{-13}\mathrm{M}_{\astrosun}$} &
  \colhead{AU} &
  \colhead{AU} &
  \colhead{dex}
}
\startdata
 11413 &  0.30 &  4423 &  0.117 &  2.30 &  1.68 & 1050 &  1.76 & 1117 & 2617 & -1.07\\
 30422 &  1.24 &  2117 &  0.053 &  1.54 &  3.09 &  502 &  1.56 & 859 & 2011 & - \\
 31295 &  0.89 &  3547 &  0.177 &  1.82 &  3.09 &  842 &  2.60 & 552 & 1294 & -0.91\\
 110411&  1.02 &  2425 &  0.065 &  1.77 &  3.33 &  576 &  1.92 & 551 & 1292 & -0.83\\
 125162&  1.25 &  2206 &  0.060 &  1.78 &  4.19 &  524 &  2.19 & 445 & 1043 & -1.35\\
 183324&  0.52 &  1712 &  0.012 &  1.83 &  1.81 &  406 &  7.38 & 881 & 2065 & -1.41\\
\enddata
\tablenotetext{a}{The uncertainties in the best-fit parameters ($r_{\mathrm{out}}$ and $\rho$) are of about 10\%, as calculated using the $\chi^2$ minimisation. $n_{\mathrm{gas}}$ corresponds to the total number gas density of the cloud, while $t_{\mathrm{acc}}$ is the calculated timescale of the star-cloud interaction.}
\tablenotetext{b}{Physical aperture radius at the distance to the star at 24 $\mu$m.}
\tablenotetext{c}{Physical aperture radius at the distance to the star at 70 $\mu$m.}
\tablenotetext{d}{Average metallicities calculated for the stars based on \cite{Heiter02}.}

\end{deluxetable*}

\subsection{Interstellar clouds and debris disks}

Our modelled mid-infrared fluxes are consistent with the {\it Spitzer} photometry of our sample of stars. \cite{Su06} have shown that the infrared photometry of these stars can also be fitted using larger (a few $\mu$m) grains distributed in debris disks, which means that simple SED fitting cannot distinguish between the two phenomena \citep[see also ][]{Kalas02}. The uncertainties and variations in the ISM-interaction models due to cloud size, photometric aperture, and $\rho_{dust}$ reinforce this point. 

In addition, there is no clear observational distinction in the photometric data between the excesses of $\lambda$ Bootis stars and those of the entire set of A-stars observed with {\it Spitzer} \citep{Su06}. The incidence of excesses is similar, as is the range of 24/70 $\mu$m color temperatures. It is improbable that all A-stars with excesses result from interactions with the ISM, but at least in the case of $\delta$ Vel\footnote{and in some cases without clear bow shock morphology such as the examples discussed by \cite{Kalas02}} an excess does result from this mechanism \citep{Gaspar08}. Thus, our modeling expands on the suggestion of \cite{Kalas02} that ISM heating can produce far infrared emission that is difficult to distinguish with existing data from a debris disk for any given system. The reflection nebulae discussed in \cite{Kalas02}, however, show no sign of dynamical interaction such as bow shocks. The typical mid infrared emission (their Table 4) is one to two orders of magnitude larger than that from our objects and the derived cloud sizes are closer to 100000~AU (their Table 5) as compared to a few 1000~AU in the interacting systems presented here. We thus conclude that the two groups of objects are physically different systems.

Heated ISM would be expected to produce an extended distribution of emission by aromatic grains. A more definitive test would be to use IFU spectroscopy with MIRI on JWST to search for aromatic emission features off the star itself (spectra centered on the star will be too dominated by the photospheric emission to see such features).

It is interesting to estimate the probable division between debris disk and heated ISM systems. According to our results, very diffuse ($n = 0.5-5$ cm$^{-3}$) and small ($r_{\mathrm{out}}$ $\sim$ 0.02 pc) interstellar clouds with total dust masses of order 10\% of a lunar mass would be enough to produce many of the observed infrared excesses. We can make a rough estimate of the time an ensemble of stars will spend in such clouds from the work of \cite{Talbot77}. They determined that a typical star like the sun has passed through about 135 clouds of $n(\rm{H}) > 100$ cm$^{-3}$ and nominal diameter 7 pc in its lifetime, which corresponds to a probability of being within such a cloud of just under 1\%.

The probability for nearby stars to be within such a cloud will be lower because of the low density of the Local Bubble. Nonetheless, it is not zero: \cite{Stan05} and \cite{Braun05} have detected tiny interstellar clouds within the Local Bubble that are closely analogous to our model cloud, with diameters of a few thousand AU and densities of 5 - 100 cm$^{-3}$. It appears that a minority but still significant number of infrared excess stars are plausibly explained by the ISM-interaction process, given the low filling factor of the local ISM. 

Further understanding of this issue requires means to identify excesses due to heated ISM dust. One approach is to identify bow shock features, which may be characteristic of star-cloud interactions at the typical values of $\sim$ 30-40 km s$^{-1}$ for $v_{\rm{rel}}$ \citep{Gaspar08}. Careful deconvolution of {\it Spitzer} imaging (G\'asp\'ar et al. 2008, in preparation) may be capable of finding additional examples. Of course, since the avoidance radii derived for our program stars are typically larger than the typical debris disk outer radius, it is possible that both mechanisms could contribute to the excess from a single star. 

Outside the Local Bubble the issue is more difficult, both because the number of ISM clouds capable of generating excesses is much larger, and because imaging with {\it Spitzer} is inadequate to identify the characteristic structure from an ISM interaction.

\subsection{Contamination of stellar atmospheres}

A star will show $\lambda$ Bootis characteristics only after its atmospheric contamination exceeds a threshold. The contamination of a stellar atmosphere will be proportional to the accretion rate from an interstellar cloud. \cite{Bondi44} solved the equations of motion for gas particles that move with uniform speed in the gravitational field of a massive object. They concluded that if a star moves across an interstellar cloud with a relative velocity $v_{\rm{rel}}$ , it accretes material up to an accretion radius given by:

\begin{equation}
\label{eq16}
r_{\rm{acc}} =\frac{\sqrt {2.5} GM_{*}}{v_{\mathrm{rel}}^2 }
\end{equation}

\noindent
This radius translates into an accretion rate (the Bondi-Hoyle accretion rate) of:

\begin{equation}
\label{eq17}
\mathop M\limits^. =\pi r_{\rm{acc}}^2 \rho _{\rm{gas}} v_{\mathrm{rel}}
\end{equation}

\noindent
where $\rho_{\rm{gas}}$ is the mass density of gas (analogous to $n_{\mathrm{gas}}$). 

This equation assumes that the interstellar gas can be treated as a fluid, i.e., the gas kinetic and potential energy is efficiently dissipated along the symmetry axis behind the star. Strictly speaking, this only happens if the mean free path is much smaller than the accretion radius, a condition that is not met under the conditions in our model clouds. The time for the star to move through the cloud is short compared to the ionization time scale and hence the gas is mostly neutral, so the mean free paths are long (although carbon and other  metals can be largely ionized by the diffuse interstellar UV radiation field). However, \cite{Bondi44} postulated an increase in the IS gas density in the wake behind the star, since many gas particle trajectories cross on the axis. Such behavior is indeed found in multi-dimensional numerical simulations of Bondi-Hoyle accretion (see \cite{Edgar04} for a summary). The higher density would facilitate accretion. In addition, the best-fit densities from our model are "averages". There is evidence for structure in the ISM down to scales of only tens of AU (e.g., \cite{Frail94}, \cite{Desh00}, \cite{Hart03}). Thus, local very-small-scale density enhancements may result in accretion at higher rates than implied by the average densities. 

The details of the accretion are likely, therefore, to be far more complex than the Bondi \& Hoyle formalism applied to the average densities. Nonetheless, we use Eqns.\ (15) and (16) with these densities to make rough estimates of accretion rates expected. We estimate the accretion radii ($r_{\rm{acc}}$) and the accretion rates ($\mathop M\limits^.$) for our sample stars, assuming they have encountered our model cloud. Since we assume that the cloud has a uniform gas density, the accretion occurs over the time required for the star to cross the cloud. For example, taking a star of 2.0 M$_\odot$, $n_{\mathrm{gas}} = 4.0$ cm$^{-3}$, $v_{rel}= 15$ km s$^{-1}$ (from Table 3), and $r_{\mathrm{out}} = 1600$ AU  (the cloud radius), we find an accretion rate of $2.3 \times 10^{-14}$ M$_\odot$ yr$^{-1}$ (see also \cite{Gaspar08}). \cite{Turcotte93} estimated that an accretion rate $\ge 10^{-14}$ M$_\odot$ yr$^{-1}$ will yield $\lambda$ Bootis characteristics in a 8000 K main-sequence star, which means that our clouds, at least for small velocities, are in the low density limit of the phenomenon.  

The time for accretion in our small clouds is short; 15 km s$^{-1}$ gives a net motion of 3 AU yr$^{-1}$. Therefore, our model cloud is traversed in only 1000 yr and the total accreted mass is $\sim$ 2 $\times$ 10$^{-11}$ M$_\odot$. Evolutionary models show that A-type stars with masses between 2.0 and 2.5 M$_\odot$ have convection zone masses ranging from 10$^{-11}$ to 10$^{-10}$ M$_\odot$. At low velocities, the accreted mass is thus comparable to the mass in the convective zone and an observable contamination is possible. As we have mentioned, two of our stars might have velocities close to this lower limit.  At larger velocities, close to our upper limit of 40 km s$^{-1}$, the accretion rates drop by two orders of magnitude, and for stars with large relative velocities it seems unlikely that such a small contamination could dominate the abundances in this zone to produce the very large anomalies (e.g., two orders of magnitude) seen in $\lambda$ Bootis stars. 

Apart from changes in $v_{\mathrm{rel}}$, other plausible alternative circumstances can produce more favorable results. For example, there may be larger clouds, as implied if the Local Bubble has a typical self-similar ISM structure. In addition, the conditions required by the ISM accretion hypothesis become much more plausible outside the Local Bubble, which in some directions only extends to $\sim$ 50 pc \citep{Lallement03}. 

In summary, we find that the requirements for adequate accreted matter to produce the $\lambda$ Bootis abundance signature are significantly more stringent than those needed to yield an infrared excess. Alternatively, this conclusion can be stated that any star that is actively accreting would produce a substantial excess. Nonetheless, this result does not predict that every $\lambda$ Bootis-type star should have an infrared excess. The accreted matter from a small IS cloud will remain in the outer layer of the star for long enough that the star is likely to have moved out of the cloud well before the $\lambda$ Bootis characteristics decay away.  

\section{Conclusion}

We have constructed a geometrical and thermal emission model for the interaction of hot stars with diffuse interstellar clouds. We find that the predicted flux emitted by dust in the clouds is consistent with {\it Spitzer} photometry of $\lambda$ Bootis stars. However, we also find that simple SED fitting is not enough to distinguish between infrared excesses generated in this manner and those due to planetary debris disks. Within the Local Bubble, we estimate that the contamination of debris disk samples will be low, but further away it may be significant. 

Our model provides order of magnitude estimates for the hypothetical cloud parameters. Small ($\sim$ 0.02 pc) and diffuse ($\sim$ 0.5 - 5 cm$^{-3}$) clouds with masses of about 10\% of a lunar mass in solid material are sufficient to produce the observed infrared excess around these stars. It may be possible to identify examples of such interactions by observing bow shocks like the one found by \cite{Gaspar08} for $\delta$ Velorum. For these cloud parameters, we estimate accretion rates of gas onto the stellar surface between 10$^{-16}$ and 10$^{-14}$ M$_{\odot}$ yr$^{-1}$, probably too low to produce $\lambda$ Bootis characteristics. However, our results suggest that ISM interactions under certain favorable conditions (low relative velocity between cloud and star, higher than average cloud density, larger clouds) may yield both the infrared excesses and abundance anomalies of such stars. These conditions may
be relatively common outside the Local Bubble.

\acknowledgements

This work is based in part on observations made with the {\it Spitzer Space Telescope}, which is operated by the Jet Propulsion Laboratory, California Institute of Technology under NASA contract 1407. Support for this work was provided by NASA through Contract Number 1255094 issued by JPL/Caltech. Part of this work was carried out at the Space Telescope Science Institute, under a Director's Discretionary Research Funds (DDRF) grant no. D0001.82352. We would like to thank Dr. Bernhard Brandl for kindly allowing one of us (JRMG) to finish this work during his PhD project.

\end{document}